\documentclass{elsart}
\usepackage{amssymb}
\usepackage{amsmath}\allowdisplaybreaks
\usepackage{graphicx}
\usepackage[normalem]{ulem}
\usepackage[dvips]{color}
\begin{document}
\begin{frontmatter}

\title{Too massive neutron stars: The role of dark matter?}
\author [XMU]{Ang Li\thanksref{info1}},
\author [XMU]{Feng Huang\thanksref{info2}}
\author [PKU]{and Ren-Xin Xu\thanksref{info3}}
\address[XMU]{Department of Physics and Institute of Theoretical Physics and Astrophysics, Xiamen University, Xiamen
361005, China}
\address[PKU]{School of Physics and State Key Laboratory of Nuclear Physics and Technology, Peking University, Beijing 100871, China}
\thanks[info1]{liang@xmu.edu.cn}
\thanks[info2]{fenghuang@xmu.edu.cn}
\thanks[info3]{r.x.xu@pku.edu.cn}

\begin{abstract}

The maximum mass of a neutron star is generally determined by the
equation of state of the star material. In this study, we take into
account dark matter particles, assumed to behave like fermions with
a free parameter to account for the interaction strength among the
particles, as a possible constituent of neutron stars. We find dark
matter inside the star would soften the equation of state more
strongly than that of hyperons, and reduce largely the maximum mass
of the star. However, the neutron star maximum mass is sensitive to
the particle mass of dark matter, and a very high neutron star mass
larger than 2 $M_{\odot}$ could be achieved when the particle mass
is small enough, being $M_{\odot}$ the mass of the sun. Such kind of
dark-matter-admixed neutron stars could explain the recent
measurement of the Shapiro delay in the radio pulsar PSR J1614-2230,
which yielded a neutron star mass of 1.97 $\pm$ 0.04 $M_{\odot}$
that may be hardly reached when hyperons are considered only, as in
the case of the microscopic Brueckner theory. Furthermore, in this
particular case, we point out that the dark matter around a neutron
star should also contribute to the mass measurement due to its pure
gravitational effect. However, our numerically calculation
illustrates that such contribution could be safely ignored because
of the usual diluted dark matter environment assumed. We conclude
that a very high mass measurement of about 2 $M_{\odot}$ requires a
really stiff equation of state in neutron stars, and find a strong
upper limit ($ \leqslant 0.64$ GeV) for the particle mass of
non-self-annihilating dark matter based on the present model.

\vspace{5mm} \noindent {\it PACS:}
26.60.Dd    
26.60.Kp    
95.35.+d    
97.60.Jd    
\end{abstract}

\begin{keyword}
Neutron stars; Equation of state; Dark matter
\end{keyword}

\end{frontmatter}


\section{Introduction}

Neutron star~(NS), a new form of compact star with degenerate
neutrons as predicted by Landau in 1932, is generally believed to
have a maximum mass, beyond which the star will be unstable and
collapse into a black hole. When considering a NS as free Fermi gas
of neutrons, the balance between the star's gravitational
self-attraction and neutron degeneracy pressure leads to the
original Oppenheimer-Volkoff mass limit of approximately 0.7
$M_{\odot}$. Incorporating the strong interaction between neutrons
will certainly increase this value because of the repulsive nature
of the short-range core. However, when hyperons are included as
another constituent of the star, a softer equation of state~(EoS)
will be obtained with a consequent reduction of the maximum NS mass.
An exact prediction for the maximum mass is difficult due to the
large uncertainty when extrapolating the EoS of dense matter from
relatively low densities in nuclear experiments to very high
densities in astrophysical objects. The final conclusion will depend
on the composition of a NS and how we describe the interactions
between its constituents.

The recent measurement \cite{Dem10} of the Shapiro delay in the
radio pulsar PSR J1614-2230 yielded a mass of 1.97 $\pm$ 0.04
$M_{\odot}$. Such a high NS mass measurement has raised great
interests in the structure and composition of NSs, since it might
rule out many predictions of non-nucleonic components (free quarks,
mesons, hyperons) in NS interiors~\cite{Hei00,Sch05,Li10k}.

For example, a large NS maximum mass larger than 2 $M_{\odot}$ is
obtained from nucleonic EoS from the microscopic Brueckner theory,
but a rather low value below 1.4 $M_{\odot}$ is found for hyperon
stars (HSs) in the same method~\cite{Bur10,Bur11}, namely so-called
hyperon puzzle. Although the present calculation did not include
three-body hyperon interaction due to the complete lack of
experimental and theoretical information, it seems difficult to
imagine that these could strongly increase the maximum mass, since
the importance of hyperon-hyperon potentials should be minor as long
as the hyperonic partial densities remain limited. However, there is
still a possibility that if there is universal strong repulsion in
all relevant channels the maximum mass may be significantly
raised~\cite{Vid11}, so the including of an improved hyperon-nucleon
and hyperon-hyperon potentials and hyperonic three-body forces is
still appealing to settle this apparent contradiction, which badly
needs further experimental data. In addition, the presence of a
strongly-interacting quark matter, in the star's interior (i.e.,
hybrid star), is proposed to be a good candidate for troubleshooting
this problem \cite{Sch11}. However, NS masses substantially above 2
$M_{\odot}$ seem to be out of reach even for hybrid stars using most
of effective quark matter EoS (bag model~\cite{Bur02}, NJL
model~\cite{Bal03}, color dielectric model~\cite{Mai04}). A hybrid
star with 2 $M_{\odot}$ is only allowed when using the
Dyson-Schwinger approach for the description of quark
matter~\cite{Che11}.

Dark matter~(DM), as another possible constituent in NS interior,
has been taken into account and a new kind of compact star, i.e.,
DM-admixed NS, has been studied recently in several articles
\cite{Gar10,Leu11,San11,Gol11,San09,Ber08,Kou08,Kou10,Lav10}. The
general effect induced by DM inside NS is complicated due to the
lack of information about the particle nature of DM. DM could
annihilate, such as the most favored candidate, neutralino, which
may lead to sizable energy deposit and then enhance the thermal
conductivity or trigger the deconfinement phase transition in the
core of NS for the emergency of a quark star, as illustrated by
Perez-Garcia et al. in \cite{Gar10}. Such quark star objects are at
present very uncertain in theory and could easily accord with
astrophysical measurements within the modification of model
parameters \cite{Li10,Li11}. Another generally considered DM
candidate is the non-self-annihilating particle, such as the newly
interesting mirror DM (\cite{mir} and references therein) or
asymmetric DM (\cite{asy} and references therein). When they
accumulate in NSs, the resulting maximum mass is then rather
sensitive to the EoS model of DM. Assuming that the DM component is
governed by an ideal Fermi gas, Leung et al. \cite{Leu11} studied
the various structures of the DM-admixed NSs by solving the
relativistic two-fluid formalism. Ciarcelluti \& Sandin \cite{San11}
approximated the EoS of mirror matter with that of ordinary nuclear
matter, varied the relative size of the DM core, and explained all
astrophysical mass measurements based on one nuclear matter EoS.
More recently, Goldman \cite{Gol11} discussed the implications of
asymmetric DM on NSs, and argued that a large mass will pose no
problem for a mixed NS. They adapted scaled EoSs of nuclear matter
for that of the dark baryons, and used two central energy densities
for the solving of NS structure equations. In this study, we will
consider non-self-annihilating DM particles as fermions, and the
repulsive interaction strength among the DM particles is assumed to
be a free parameter $m_\mathcal{I}$ as in \cite{Nar06}. Different to
previous DM-admixed NS models,  we take the total pressure (energy)
density as the simple sum of the DM pressure(energy) and NS pressure
(energy), the general dependence of the mass limit on DM particle
mass and the interaction strength is then presented based on the
present model.

In addition, the non-self-annihilating DM, mirror DM or asymmetric
DM, is generally believed to simply accumulate during the whole
evolution series from the proton-star to the final compact state. As
the heavy DM particles usually do not collapse with the ordinary
matter, an extended halo around the star is
formed~\cite{Ste78,Pre85}, therefore there should exist an extra
general-relativistic mass effect from the halo. This is particularly
relevant for the mass measurement of PSR J1614-2230, because the
inferred large NS mass is based on the large Shapiro delay and
Keplerian orbital parameters of a binary system \cite{Dem10}, and
information on the size of the NS in the binary system is actually
not clear. It may be possible that the inferred mass comprises the
mass of the star and also the mass of a possible DM halo. It is in
the present article that for the first time the mass contribution
from the possible extended halo is taken into account. For that we
should consider carefully the spatial scale of the related halo and
the DM density around the position of the binary system (see the
following section for details).

This paper is arranged as follows. The details of our theoretical
model are presented in \S2, followed by the numerical results.
Conclusions and discussions are given in \S3.

\section{The model}
\label{sect:model}

DM particles, as the most abundant matter component in the universe,
could accrete onto stars due to their kinetic energy loss in the
scattering process and also gravitationally trapped inside or around
the star during the whole star evolution stage. DM particles being
scattered inside the star would modify the local pressure-energy
density relationship of the matter and hence change the theoretical
prediction of the gravitational mass of the star. DM particles left
behind the star could form an extended halo around the star, which
is expected to increase the measured mass of the star. We will study
in detail these two aspects in the following two subsections
respectively.

\subsection{DM-admixed NS model}

The structure equations for compact stars, namely the Einstein
field equations for hydrostatic equilibrium (i.e, the
Tolman-Oppenheimer-Volkov (TOV) equations) are written as:
\begin{equation}
 \frac{dP(r)}{dr}=-\frac{Gm(r)\mathcal{E}(r)}{r^{2}}
 \frac{\Big[1+\frac{P(r)}{\mathcal{E}(r)}\Big]
 \Big[1+\frac{4\pi r^{3}P(r)}{m(r)}\Big]}
 {1-\frac{2Gm(r)}{r}},
   \label{tov1:eps}
\end{equation}
\begin{equation}
\frac{dm(r)}{dr}=4\pi r^{2}\mathcal{E}(r)
  \label{tov2}
\end{equation}
being $G$ the gravitational constant. $P$ and $\mathcal{E}$ denote
the pressure and energy density. The EoS of the star, relating $P$
and $\mathcal{E}$, is needed to solve the above set of equations. In
our DM-admixed NS model, $P = P_N + P_{\chi}, \mathcal{E} =
\mathcal{E}_N +\mathcal{E}_{\chi}$, with the subscript $N(\chi)$
representing NS matter (DM).

The EoS of the ordinary NS matter is handled in the following way:
(i) We treat the interior of the stars as $\beta$-equilibrium
nuclear matter (corresponding to NSs) or hypernuclear matter
(corresponding to HSs), with certain amount of leptons to maintain
charge neutrality. The hadronic energy density we use in the article
is based on the microscopic parameter-free Brueckner-Hartree-Fock
nuclear many-body approach, employing the latest derivation of
nucleon-nucleon microscopic three-body force~\cite{Li08}. When
performing the study of HSs, the very recent Nijmegen extended
soft-core ESC08b hyperon-nucleon potentials~\cite{Sch11} is included
as well. The EoS can be computed straightforwardly after adding the
contributions of the noninteracting leptons~\cite{Sch11}. (ii) For
the description of the NS/HS crust, we join the hadronic EoS with
those by Negele and Vautherin~\cite{nv} in the medium-density
regime, and those by Feynman-Metropolis-Teller~\cite{fmt} and
Baym-Pethick-Sutherland~\cite{bps} for the outer crust.

The DM part may stabilize itself in a barotropic state in the same
way as in the case of ordinary matter, but it is very difficult to
determine what is the EoS of DM. We will take DM as Fermi gas with
$m_\mathcal{I}$ accounting for the energy scale of the interaction,
and write the energy density and pressure of DM as those of a
self-interacting Fermi gas as \cite{Nar06}:
\begin{eqnarray}
\mathcal{E_{\chi}} &=& \frac{m_{\chi}^4}{ \pi ^2} \int_0^{k_F/m_{\chi}}
x^2 \sqrt{1+x^2} dx + \left(\frac{1}{ 3 \pi ^2}\right)^2 \frac{k_F^6}{ m_\mathcal{I}^2}
\\
P_{\chi} &=& \frac{m_{\chi}^4}{3 \pi^2} \int_0^{k_F/m_{\chi}}
\frac{x^4}{\sqrt{1+x^2}} dx + \left(\frac{1}{3 \pi ^2}\right)^2
\frac{k_F^6}{m_\mathcal{I}^2}
\label{eqi:eos}
\end{eqnarray}
where $m_\chi$ is the mass of DM particles, and the Fermi momentum
$k_F$ is related to the number density $\rho$ by $k_F =
(3\pi^2\rho)^{1/3}$ .

For weak interaction (WI) the scale $m_\mathcal{I}$ can be
interpreted as the expected masses of W or Z bosons generated by the
Higgs field, which is $\sim300$ GeV. For strongly interacting (SI)
DM particles, $m_\mathcal{I}$ is assumed to be $\sim100$ MeV,
according to the gauge theory of the strong interactions. This is a
wide enough range of energy scale, and we hope the calculation would
cover most of the promising DM candidates.

As far as the pressure and energy density of NS and DM have been
determined, we then start with a central mass density $\mathcal{E}(r
= 0)$, and integrate out until the surface density equals that of
iron. This gives the stellar radius $R$ and its enclosed mass $M =
m(R)$. Each EoS is related to a NS equilibrium sequence with
different central mass density, and there is a maximum value of
central density (or central pressure) for each EoS, which
corresponds to the maximum weight of the star sequence. The mass of
the stars can not be larger than the maximum mass value because it
will unavoidably collapse due to unbearable gravity. If a
theoretical model predicts a maximum value of NS which is lower than
the mass measurements of pulsars in the market, we say the model
fails to explain the experiments and is ready to be improved or
rejected.

\begin{figure}
\begin{center}
\includegraphics[width=13.0cm]{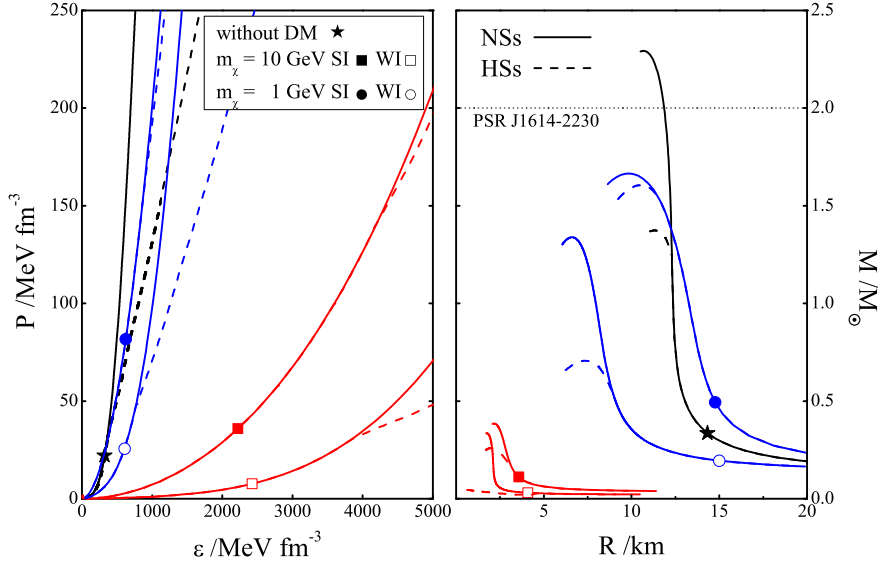}
\end{center}
\caption{(Color online) Equations of state (left panel) and
mass-radius relations (right panel) of the DM-admixed NSs (solid
curves) and HSs (dashed curves) with a recently-determined DM
particle mass $m_{\chi}$ = 10 GeV for SI and WI DM, to be compared
with the case without DM. The results with a modified DM particle
mass with $m_{\chi}$ = 1 GeV are also shown. The $\sim 2$
$M_{\odot}$ limit of PSR J1614-2230 is indicated with a horizontal
line. } \label{fig2}
\end{figure}

Fig.~1 presents EoSs (left panel) and mass-radius relations (right
panel) of the DM-admixed NSs (solid curves) and HSs (dashed curves)
with a recently-determined DM particle mass $m_{\chi}$ = 10
GeV~\cite{Buc11} for SI and WI DM, to be compared with the case
without DM. The mass of 10 GeV accounts for a consistent description
about various recent direct detection experiments, with which the
EoSs are substantially softened after the inclusion of DM
contribution both in SI and WI cases. This leads to smaller maximum
masses, as shown in the right panel. A maximum value of 2.29
$M_{\odot}$ (1.37 $M_{\odot}$) when DM is not included is decreased
to 0.39 $M_{\odot}$ (0.26 $M_{\odot}$) in the SI case, and to 0.34
$M_{\odot}$ (0.05 $M_{\odot}$) in the WI case for NSs (HSs), where
the recent-observed $\sim 2 M_{\odot}$ mass measurement is indicated
with a dotted horizontal line. The softening of DM in this case is
seen to be quite evident, even stronger than that of hyperons.
However, current predictions of the DM particle mass span the range
from keV as the sterile neutrino to around TeV as weakly interacting
massive particles (usually shortened as WIMP). If we use a decreased
mass of $m_{\chi}$ = 1 GeV to redo the calculation, the evident
softening effect of DM is somehow weakened as illustrated in the
same figure, and a larger maximum masses are obtained, namely 1.67
$M_{\odot}$ (1.61 $M_{\odot}$) in the SI case, and 1.34 $M_{\odot}$
(0.71 $M_{\odot}$) in the WI case for NSs (HSs). This demonstrates
an interesting sensitive dependency of the maximum mass on the DM
particle mass $m_{\chi}$, which needs further exploration.

\begin{table}
\begin{center}
\begin{minipage}{135mm}
\caption{\small Characteristics of the maximum mass configurations
(maximum masses $M$, corresponding radii $R$ and central number densities $\rho_c$)
for different DM mass $m_{\chi}$ and composition.}
\begin{tabular}{c|c|cccccc}
\hline  \hline
$m_{\chi}$ (GeV)& &\multicolumn{3}{c}{SI} &\multicolumn{3}{c}{WI}\\
&    & $M (M_{\odot})$& $R$ (km) &$\rho_c$ (fm$^{-3}$)& $M (M_{\odot})$ & $R$ (km) & $\rho_c$ (fm$^{-3}$)\\
\hline
0.01  & NS & 2.96 & 17.3 & 0.35 & 2.11 & 12.4 & 0.77  \\
      & HS & 2.96 & 17.3 & 0.35 & 2.11 & 12.4 & 0.77  \\
\hline
0.1   & NS & 2.88 & 16.8 & 0.36 & 2.06 & 11.7 & 0.82  \\
      & HS & 2.88 & 16.8 & 0.36 & 2.06 & 11.7 & 0.82  \\
\hline
   1  & NS & 1.67 & 9.85 & 0.68 & 1.34 & 6.61 & 1.39  \\
      & HS & 1.61 & 10.5 & 0.61 & 0.71 & 7.39 & 1.32  \\
\hline
10    & NS & 0.39 & 2.16 & 2.61 & 0.34 & 1.74 & 4.12  \\
      & HS & 0.26 & 1.99 & 3.62 & 0.05 & 0.65 & 40.9  \\
\hline  \hline
\end{tabular}
\end{minipage}
\end{center}
\end{table}

\begin{figure}
\begin{center}
\includegraphics[width=13.0cm]{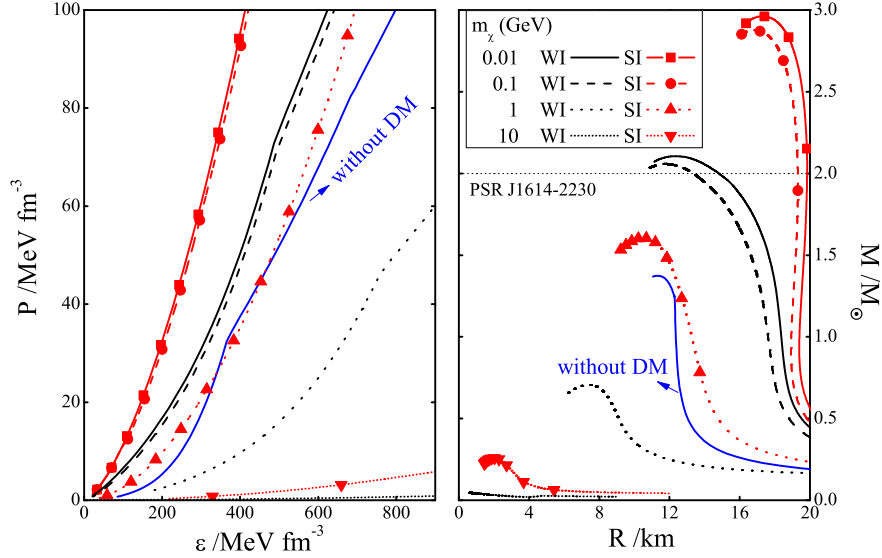}
\end{center}
\caption{(Color online) Equations of state (left panel) and
mass-radius relations (right panel) of the DM-mixed HSs with DM mass
$m_{\chi}$ ranging from 0.01GeV to 10GeV for SI (curves with symbol)
and WI (curves without symbol) DM, to be compared with the case
without DM. The $\sim 2 M_{\odot}$ limit of PSR J1614-2230 is
indicated with a horizontal line.} \label{fig2}
\end{figure}

In Table 1 we collect the calculated characteristics of the maximum
mass configurations (maximum masses, corresponding radii and central
number densities) with different DM mass $m_{\chi}$ and composition.
Because of the conflict mentioned above between the HS theoretical
model and the recent observed large mass, special attention is paid
to the HS results, which are presented in Fig.~2 of the EoSs (left
panel) and mass-radius relations (right panel) using DM mass
$m_{\chi}$ ranging from 0.01GeV to 10GeV for SI (curves with symbol)
and WI (curves without symbol) DM, to be compared with the case
without DM. It is clear that the smaller the mass of DM, the larger
the mass of the compact star could reach. If the newly measurement
of 1.97 $\pm$ 0.04 $M_{\odot}$ is required for a HS, an upper limit
on the DM mass around 0.64 GeV (0.16 GeV) are set for SI (WI) DM.

Our predication on the upper limit of DM mass could be relaxed. For
example, If a part of the measured 2 $M_{\odot}$ is deposited around
the NS, eg. 1.61 $M_{\odot}$, the upper limit of DM mass could be
increased to as large as 10 GeV. This is the reason why we further
consider the DM extended halo contribution.

\subsection{DM halo around NS}

\begin{figure}
  \vspace*{174pt}
\begin{center}
\includegraphics[width=13.8cm]{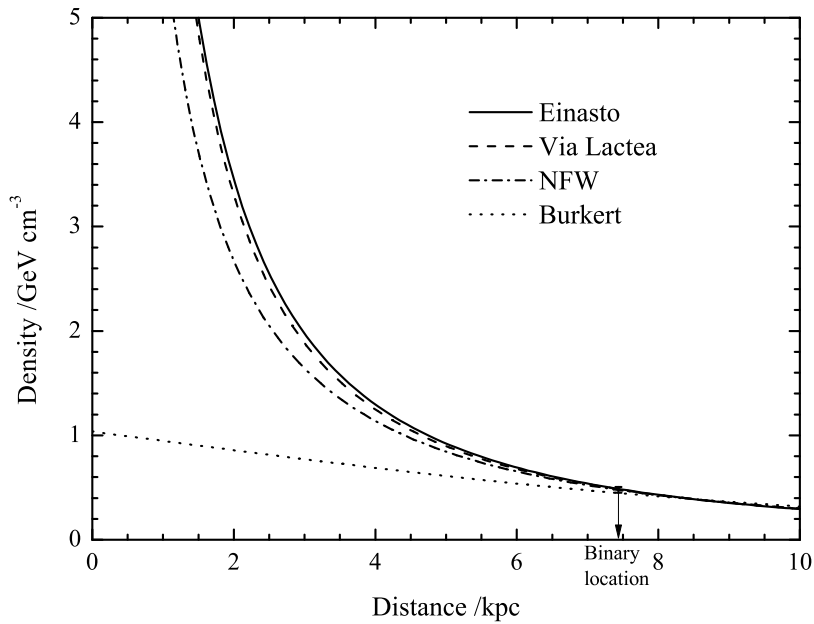}
\caption{Four models of the galactic DM density profiles employed in
the paper. $Einasto$ stands for the best-fitting Einasto density
profile obtained from the results of the Aquarius simulation
\cite{Spr08}. $Via~Lactea$ stands for a profile extrapolated from
the Via Lactea II simulation \cite{Spr08}. $NFW$ stands for the
prototypical Navarro-Frenk-White density profile \cite{Nav97}.
$Burkert$ stands for the Burker profile \cite{Bur95,Sal00}. The
location of the binary system $r$ = 7.44 kpc is indicated with an
arrow.} \label{fig3}
\end{center}
\end{figure}

To get the mass contribution of DM halo via gravitational capture,
we first should calculate the spatial scale of the related halo and
the DM density at the NS location. We estimate the size of the halo
as big as that of the possible Roche lobe of the centered PSR
J1614-2230, which is calculated using the following theoretical
formula by Eggleton~\cite{Egg}:
\begin{equation}
R = \frac{0.49 (M_1/M_2)^{2/3}} {0.6 (M_1/M_2)^{2/3} + ln[1+(M_1/M_2)^{1/3}]}~a  \label{lobe}
\end{equation}
where $a$ is the major semi-axis of this binary system which is
3$\times$10$^{11}$ cm. $M_1$ is the gravitational mass of the NS,
and $M_2$ is that of its companion star, a 0.5 $M_{\odot}$ white dwarf
(WD) \cite{Dem10}. The gravitational mass of the NS is ready to
change when incorporating the DM (as shown below), but the value of
the WD, i.e., 0.5 $M_{\odot}$, is fixed since it is implied by the detected
Shapiro delay of PSR J1614-2230 by the WD. A possible DM halo
around the WD has no influence on this value, because the
measurement is done for a complete period of the binary system.

The DM density in the extended halo around the NS is highly
dependent on the local distribution of DM density which should be
determined from the accreting history in the binary system. Here,
for a simple calculation, we adopt the density value determined by
our Galaxy density profile. We restrict our evaluation to several
spherically symmetric Galactic DM profiles, and scale the profiles
with a fixed value of 0.389 GeV/cm$^3$ at the solar position. As
shown in Fig. 2, $Einasto$ stands for the best-fitting Einasto
density profile obtained from the results of the Aquarius simulation
\cite{Spr08}. $Via~Lactea$ stands for a profile extrapolated from
the Via Lactea II simulation \cite{Spr08}. $NFW$ stands for the
prototypical Navarro-Frenk-White density profile \cite{Nav97}. The
last $Burkert$ profile is characterized by a very smooth central
cusp \cite{Bur95,Sal00}. From the celestial coordinates of PSR
J1614-2230 (16 hr 14 min right ascension and -22 degrees 30 minutes
declination) and its distance from the sun (1.2 kpc) \cite{Dem10},
we can calculate its distance from the galactic center, which is
7.44 kpc. Then the local DM densities $\rho_{\chi}$ can be evaluated
corresponding to the four profiles above, namely 0.4868 GeV/cm$^3$,
0.4832 GeV/cm$^3$, 0.4771 GeV/cm$^3$, 0.4472 GeV/cm$^3$,
respectively. Since they do not differ much, it is proper to take an
average value of $\bar{\rho_{\chi}}$=0.474 GeV/cm$^3$ for the
following calculation. Hence the contributed mass of gravitationally
captured DM particles can be finally written as
\begin{equation}
M_{\chi}= \frac{4}{3}\pi R^3 \bar{\rho_{\chi}} \label{gravitation}
\end{equation}
where we have neglected the size of the star ($\sim10$ km) compared
to its large Roche lobe ($\sim10^6$ km), and have regarded the halo
as an ideal spherical object. In this case, the extra mass
measurement contribution from the above extended halo is around
10$^{-24}$ $M_{\odot}$, which could be safely ignored. However, the
capture of DM may be further enhanced by the motion of the NSs in
close binaries \cite{Bra11}. Our adopted value should be considered
as the lower limit mass contribution. Even though,  DM mass
contribution from the extended halo alone is hard to account for the
large mass measurement even the density could be increased by
several orders in some exotic mechanism, such as some abnormal
stellar merge events of DM stars, or abnormally efficient absorbing
of DM.

 \begin{figure}
 \begin{center}
 \includegraphics[width=13.0cm]{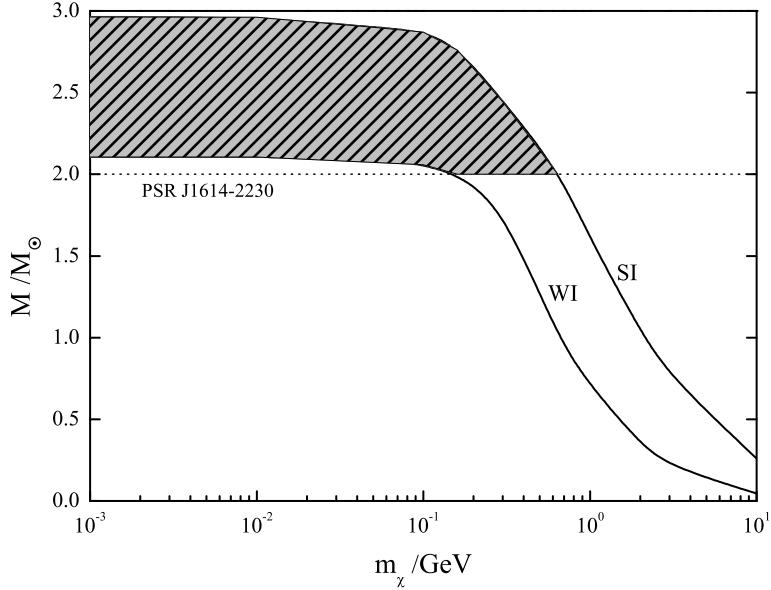}
 \end{center}
 \caption{HSs' maximum masses as a function of the particle mass of DM candidates $m_{\chi}$.
 The upper line is for $m_{\mathcal{I}}$ = 100 MeV (SI case),
 and the lower line for $m_{\mathcal{I}}$ = 300 GeV (WI case).
 Again the $\sim 2$ $M_{\odot}$ limit of PSR J1614-2230 is indicated
 with a horizontal line.} \label{fig3.eps}
 \end{figure}

Finally we summarize in Fig.~4 with a shallowed area the mass limit
of DM particle $m_{\chi}$ based on our present model, assuming the
observed PSR J1614-2230 is a HS. The $\sim 2$ $M_{\odot}$ limit is
again indicated with a horizontal line. The upper line corresponds
to $m_{\mathcal{I}}$ = 100 MeV (SI case), and the lower line to
$m_{\mathcal{I}}$ = 300 GeV (WI case). The dependence of the maximum
mass on the DM particle mass is very sensitive when the mass is
relatively large (above 0.1 GeV). For small mass value less than
0.01 GeV, the calculated mass-radius curves are very close to each
other in our model, and almost fixed results are obtained for the
maximum mass ($\sim2.96 M_{\odot}$) and corresponding radius
($\sim17.3$ km), as shown in Table 1. This is because that the
$\mathcal{E} (P)$ relation of DM has a weaker dependence on the
change of the particle mass $m_{\chi}$ based on the present
Fermi-gas model~(Equ.(\ref{eqi:eos})) when the particle mass is
small, as a result the maximum mass never exceeds 3$M_{\odot}$ and
comfortably lies below the usual constrain for NSs' mass. Moreover,
as discussed before, this mass limit from the compact star can be
referred as an upper limit for the mass of non-self-annihilating DM
particles, namely, it should obey $m_{\chi} \leqslant 0.64$ GeV.
More information on the interaction properties among DM particles
will certainly further narrow this region.

\section{Conclusion}

In this paper, we present a consistent DM-admixed NS model to
investigate the possible influence of DM on the NS mass measurement.
We take DM as Fermi-gas with certain repulsive interaction among the
DM particles and none-interaction between DM and ordinary matter as
is generally assumed. The pressure (energy density) of DM particles
scattered into the compact star could be regarded as an extra
component to the total pressure (energy density) in the TOV
equations. In this scenario, the DM ingredient is expected to soften
the total EoS and result in a reduced maximum mass. However, the
final results are sensitive to the adopted DM particle mass. The
smaller the DM particle mass, the harder the EoS or the larger the
maximum mass. The observed very massive NS requires a very stiff EoS
and then sets a strong upper limit on the DM particle mass. In our
numerical calculation, DM particle mass should less than 0.64~GeV
for SI DM and 0.16 GeV for WI DM. In order to relax such strong
constraint, we further consider the possible extended DM halo
contribution to the particular mass measurement in \cite{Dem10}.
However, due to the diluted DM environment, such kind of
contribution could be safely ignored. Some exotic mechanism, such as
abnormal stellar merge events of DM stars, or abnormally efficient
absorbing of DM, may lead to an unusual dense DM halo and then relax
the upper limit greatly. Generally, the EoS of the pulsar should be
really stiff unless there is a very dense DM halo around the compact
object. Very recently, such a high mass NS has been successfully
explained as a hybrid star described by a very stiff nucleonic
EoS~\cite{Che11} in the Brueckner theory, which is consistent with
our findings. This conclusion would be meaningful for the research
of microscopic physics. Since our present calculation is based on
the ordinary NS structure equations, we can not provide the specific
configuration of the DM-admixed NS. If one notice the quite small
values of NS radii in Table 1 for large DM particle mass, they are
more like DM-stars rather than NSs. More proper scheme should be
applied to solve the two-fluid equations with a updated reasonable
DM EoS, which is referred to a future work.

\section{Acknowledgments}
We would like to thank an anonymous referee for
many medicinal comments and suggestions,
and we acknowledge Dr. Tong Liu for beneficial
discussion. This work is funded by the National Basic Research
Program of China (Grant No 2009CB824800), the National Natural
Science Foundation of China (Grant No 10905048, 10973002, 10935001,
11078015), and the John Templeton Foundation.


\begin{thebibliography}{00}

\bibitem{Dem10}
P. Demorest, T.Pennucci, S. Ransom, M. Roberts, J. Hessels, Nature 467 (2010) 1081.

\bibitem{Hei00}
H. Heiselberg, M. Hjorth-Jensen, Phys. Rep. 328 (2000) 237.

\bibitem{Sch05}
J. Schaffner-Bielich, J. Phys. G: Nucl. Part. Phys 31 (2005) S651.

\bibitem{Li10k}
A. Li, X. R. Zhou, G. F. Burgio, H.-J. Schulze, Phys. Rev. C 81 (2010) 025806.

\bibitem{Bur10}
G. F. Burgio, H.-J. Schulze, A\&A 518 (2010) A17.

\bibitem{Bur11}
G. F. Burgio, H.-J. Schulze, A. Li, Phys. Rev. C 83 (2011) 025804.

\bibitem{Vid11}
I. Vid$\tilde{a}$na, D. Logoteta, C. Provid$\hat{e}$ncia, A. Polls, and I. Bombaci,
Europhys. Lett. 94 (2011) 11002.

\bibitem{Sch11}
H.-J. Schulze, T. Rijken, Phys. Rev. C 84 (2011) 035801.

\bibitem{Bur02}
 G. F. Burgio, M. Baldo, P. K. Sahu, and H.-J. Schulze, Phys.
Rev. C 66 (2002) 025802.

\bibitem{Bal03}
 M. Baldo, M. Buballa, G. F. Burgio, F. Neumann, M. Oertel, and
H.-J. Schulze, Phys. Lett. B 562 (2003) 153.

\bibitem{Mai04}
 C. Maieron, M. Baldo, G. F. Burgio, and H.-J. Schulze, Phys.
Rev. D 70 (2004) 043010.

\bibitem{Che11}
H. Chen, M. Baldo, G. F. Burgio, H.-J. Schulze,
Phys. Rev. D 84 (2011)  105023.

\bibitem{Gar10}
M. A. Perez-Garcia, J. Silk, J. R. Stone, Phys. Rev. Lett. 105
(2010) 141101.

\bibitem{Leu11}
S.-C. Leung, M.-C. Chu, L.-M. Lin, Phys. Rev. D 84 (2011) 107301.

\bibitem{San11}
P. Ciarcelluti, F. Sandin, Phys. Lett. B 695 (2011) 19.

\bibitem{Gol11}
I. Goldman, APPB 42 (2011) 2203.

\bibitem{San09}
F. Sandin, P. Ciarcelluti, Astroparticle Phys. 32 (2009) 278.

\bibitem{Ber08}
G. Bertone, M. Fairbairn, Phys. Rev. D, 77 (2008) 043515.

\bibitem{Kou08}
C. Kouvaris, Phys. Rev. D 77 (2008) 023006.

\bibitem{Kou10}
C. Kouvaris, P. Tinyakov, Phys. Rev. D 82 (2010) 063531.

\bibitem{Lav10}
A. de Lavallaz, M. Fairbairn, Phys. Rev. D 81 (2010) 123521.

\bibitem{Li10}
A. Li, R. X. Xu, J. F. Lu, MNRAS 402 (2010) 2715.

\bibitem{Li11}
A. Li, G. X. Peng, J. F. Lu, RAA 11 (2011) 482.

\bibitem{mir}
L. B. Okun, Mirror particles and mirror matter: 50 years of
speculations and search, Phys. Usp. 50 (2007) 380¨C389.
hep-ph/0606202.

\bibitem{asy}
H. An, S.-L. Chen, R. N. Mohapatra, and Y. Zhang JHEP, 3, (2010)
124.

\bibitem{Nar06}
G. Narain, J. Schaffner-Bielich, I. N. Mishustin, Phys. Rev. D 74
(2006) 063003.

\bibitem{Ste78}
G. Steigman, C. L. Sarazin, H. Quintana, J. Faulkner, ApJ 83 (1978)
1050.

\bibitem{Pre85}
W. H. Press, D. N. Spergel, ApJ 296 (1985) 679.

\bibitem{Li08}
Z. H. Li, U. Lombardo, H.-J. Schulze, W. Zuo, Phys. Rev. C 77 (2008)
034316.

\bibitem{nv}
J. W. Negele, D. Vautherin, Nucl. Phys. A 207 (1973) 298.

\bibitem{fmt}
R. Feynman, F. Metropolis, E. Teller, Phys. Rev. C 75 (1949) 1561.

\bibitem{bps}
G. Baym, C. Pethick, D. Sutherland, ApJ 170 (1971) 299.

\bibitem{Buc11}
M. R. Buckley, D. Hooper, T. M. P. Tait, Phys. Lett. B 702 (2011)
216.

\bibitem{Egg}
P. P. Eggleton, ApJ 268 (1983) 368.

\bibitem{Spr08}
V. Springel, J. Wang, M. Vogelsberger, A. Ludlow, A. Jenkins, A.
Helmi, J. F. Navarro, C. S. Frenk, S. D. M. White, MNRAS 391 (2008)
1685.

\bibitem{Nav97}
J. F. Navarro, C. S. Frenk, S. D. M. White, Astrophys. J. 490 (1997)
493.

\bibitem{Bur95}
A. Burkert, ApJL 447 (1995) L25.

\bibitem{Sal00}
P. Salucci, A. Burkert, ApJL 537 (2000) L9.

\bibitem{Bra11}
L. Brayeur, P. Tinyakov, arXiv:1111.3205.


\end{thebibliography}
\end{document}